\documentclass[aps,prl,twocolumn,nofootinbib]{revtex4}
\usepackage{graphicx}

\begin{document}
\preprint{hep-th/0605263, UCB-PTH-06/10, LBNL-60251}
\title{Holographic probabilities in eternal inflation}

\author{Raphael Bousso} 
\affiliation{Center for Theoretical Physics, Department of Physics,
University of California, Berkeley, CA 94720-7300, U.S.A.; \\
Lawrence Berkeley National Laboratory, 
Berkeley, CA 94720-8162, U.S.A.}
\begin{abstract}
  In the global description of eternal inflation, probabilities for
  vacua are notoriously ambiguous.  The local point of view is
  preferred by holography and naturally picks out a simple probability
  measure.  It is insensitive to large expansion factors or lifetimes,
  and so resolves a recently noted paradox.  Any cosmological measure
  must be complemented with the probability for observers to emerge in
  a given vacuum.  In lieu of anthropic criteria, I propose to
  estimate this by the entropy that can be produced in a local patch.
  This allows for prior-free predictions.
\end{abstract}

                    
\maketitle

The evidence for nonvanishing vacuum energy suggests that fundamental
theory has an enormous number of long-lived, metastable
vacua~\cite{Wei87}.  Happily, string theory appears to satisfy this
criterion~\cite{BP,KKLT}.  Then many low-energy parameters will not be
determined uniquely, but statistically.  To make pre- or
post-dictions, one must survey representative samples of the landscape
of vacua~\cite{Sus03,DenDou04b}.  But this is not enough.
Cosmological dynamics may favor the production of some vacua and
suppress others.  The predictivity of fundamental theory hinges on a
quantitative understanding of this effect.

The vacua with positive cosmological constant trap the universe in
eternal inflation.  They decay only locally, by producing bubbles of
new vacua.  Thus, every vacuum in the landscape will be realized an
infinite number of times in different, causally disconnected regions.
Each bubble expands to become an infinite open universe embedded in
the global spacetime.

To regulate these infinities, one might compare the prevalence of
different vacua at finite time, and then take a late time limit.  But
this task is plagued by ambiguities~\cite{LinLin94}.  Should vacuum
$i$ be weighted by the number of $i$-bubbles, or by their volume?
Worse, both quantities depend on the choice of time variable, and
there is no preferred time slicing in eternal inflation.  A number of
interesting probability measures have been proposed, most recently in
Refs.~\cite{GarVil01,GarSch05,EasLim05}.  That they give different
answers illustrates the intricacy of the problem.  One can imagine
other prescriptions, and practically any answer can be obtained by
devising a suitable time slicing.


Here I will develop a probability measure by appealing only to a
single causally connected region, or causal diamond~\cite{Bou00a}.
This is called the local, or causal, or holographic point of view.  My
approach will encounter none of the ambiguities listed above.
Moreover, there are independent reasons to embrace this viewpoint:
From the quantum properties of black holes, we have learned that the
simultaneous description of two causally disconnected regions leads to
paradoxes, which are resolved if we stick to describing only what any
one observer can measure~\cite{SusTho93}.  In fact, the situation in
eternal inflation is worse than for black holes.  An observer outside
a black hole can compute the interior geometry from initial
conditions, but an observer in eternal inflation cannot predict when
and where bubbles will form, and so cannot distinguish between
macroscopically distinct global metrics~\cite{BFY1}.  Thus, the local
observer cannot even construct a global geometry about whose slicing
one could argue.

I consider only a single worldline, so the task breaks up into 1.\
({\em Prior probability}) How likely is it for the worldline to enter
vacuum $i$?  2.\ ({\em Weighting}) What is the probability that
observers will emerge in vacuum $i$?  On the latter issue, I will find
that the causal viewpoint permits the elimination of anthropic
selection criteria---which are hard to specify for widely varying low
energy theories---in favor of prior-free thermodynamic conditions for
the emergence of complex phenomena such as observers.

\paragraph{Prior probability}
Consider a landscape with vacua $i$.  These should include metastable
vacua, which eventually decay into other vacua.  There may also be
``terminal'' vacua, which do not decay (typically, the vacua with
nonpositive cosmological constant).  If so, the landscape is called
terminal; otherwise, ``cyclic''.  (There is empirical evidence that
the landscape is terminal~\cite{DysKle02,BouFre06}.  Moreover, the
landscape of string theory is terminal.)  There will be no need to
restrict to the terminal case here, but I will assume that the
landscape is connected: every vacuum can be reached from any
meta\-stable vacuum by some sequence of decays.

How likely is it for the worldline to enter a given vacuum on its way
through the landscape?\footnote{The question is {\em not\/} how much
  time the worldline is likely to spend in $i$.  Complex phenomena
  such as observers arise between bubble formation and thermalization.
  Typical lifetimes of vacua are exponentially longer than this
  out-of-equilibrium period, so their inclusion at this point would
  lead to huge correction factors in the final section.  The length of
  the prehistory is equally irrelevant.  Never mind how long the
  worldline lingers in a meta\-stable vacuum $a$; the question is
  which vacuum it enters next.}  Let $\kappa_{ij}$ be the probability
per unit proper time for a geodesic worldline in vacuum $j$ to enter
vacuum $i$.  Normalize each column of $\kappa$ to sum to 1, $\eta_{ia}
= \kappa_{ia}/\sum_j\kappa_{ja}$, except for columns corresponding to
terminal vacua, which vanish.  The matrix $\eta$ describes the {\em
  relative\/} probability to decay from $a$ to $i$.

Now, draw a root node labeled $o$, corresponding to the initial vacuum
the worldline starts out in.  For each vacuum $i$ that $o$ can decay
into, draw a branch connecting $o$ with a new node labeled with the
new vacuum $i$.  Next to each branch, write the relative probability,
$\eta_{\rm branch}$, for this decay channel (in this case, $\eta_{\rm
  branch}=\eta_{io}$).  Then repeat this procedure for each metastable
new vacuum.  This will generate a tree.

Next, compute a raw (i.e., unnormalized) probability for each vacuum
in the landscape.  For each path from the root node to the vacuum in
question, multiply the branch probabilities; then sum up the results:
\begin{equation}
  P_i = 
  \sum_{\begin{array}{c} \rm all~nodes\\ {\rm labeled~} i\end{array}}~ 
  \prod_{\begin{array}{c} 
      \rm the~branches~connecting\\ \rm the~root~to~the~node \end{array}} 
  \!\!\!\!\!\eta_{\rm branch}
\label{eq-raw}
\end{equation}
The normalized probability for a worldline to pass through vacuum $i$
is $p_i = P_i/\sum_j P_j$.

\begin{figure}
\includegraphics[width=8.5cm]{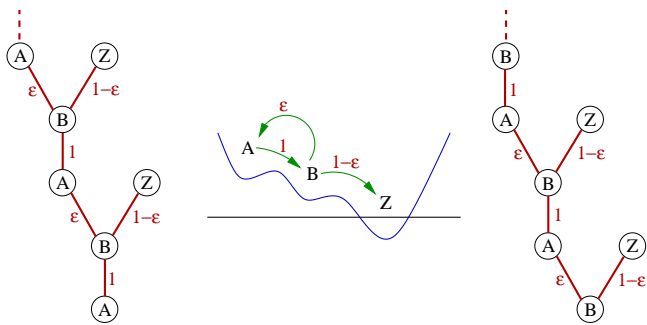}
\caption{\label{fig-abz}
A landscape with two metastable vacua and one terminal
  vacuum.  The tree on the left corresponds to a worldline starting in
  vacuum $A$ (the initial vacuum, or root).  The tree on the right
  starts with vacuum $B$.  The unnormalized probability for vacuum $i$
  is obtained by computing the probability for each path leading up
  from the root to $i$ (the product of the numbers along the path),
  and summing over all paths.}
\end{figure}

For a simple example, consider a landscape with two metastable vacua
$A$ and $B$, and a terminal vacuum $Z$, as shown in
Fig.~\ref{fig-abz}.
In this model, $A$ can only decay to $B$ ($\eta_{BA}=1$).  The vacuum
$B$ decays to $Z$ with probability $\eta_{ZB} = 1-\epsilon$, or back
up to $A$ with probability $\eta_{AB}=\epsilon$.

First, suppose that the initial vacuum is $A$.  From the associated
tree (Fig.~\ref{fig-abz}, left), one sees that there are infinitely
many paths leading into each vacuum.  For vacuum $A$, the paths are
are $ABA,~ABABA,~\ldots$, giving a raw probability $P_A = \epsilon +
\epsilon^2 + \ldots = \epsilon/(1-\epsilon)$.  For vacuum $B$, the
paths are $AB$, $ABAB$, etc., and the vacuum $Z$ arises from paths
$ABZ$, $ABABZ$, etc.  After normalization one obtains $p_A =
\epsilon/2$, $p_B = 1/2$, $p_Z = (1-\epsilon)/2$.

Now, suppose that the initial vacuum is $B$ (Fig.~\ref{fig-abz},
right).  One finds $p_A = p_B = \epsilon/(1+\epsilon)$, $p_Z =
(1-\epsilon)/(1+\epsilon)$.  As one would expect for a single
worldline, the probability to pass through a given vacuum can depend
on the initial vacuum.\footnote{Some of the extant proposals depend
  strongly on initial conditions~\cite{GarVil01}, others more
  weakly~\cite{GarSch05}.  But this is hardly a criterion for
  evaluating them, since we have neither observational nor theoretical
  grounds to demand {\em a priori\/} that the result of this
  particular dynamical process be insensitive to the starting point.
  The initial probability distribution is an independent theoretical
  problem; see Refs.~\cite{Vil98b} for an opinionated discussion of
  some proposals.}  It is interesting to take note of the limiting
values of the above probabilities as $\epsilon \rightarrow 0$, or
$\epsilon \rightarrow 1$.  These are physically the most relevant
cases, because the rates for different decay channels generically
differ by exponentially large factors.

The formulation so far is not quite perfect, since the raw
probabilities in Eq.~(\ref{eq-raw}) need not be finite.  It is useful
to think of the tree in terms of a conserved probability current,
which enters at the root (the source) and flows up, ending up
exclusively in terminal vacua (the sinks).  It follows that the total
{\em raw\/} probability for all terminal vacua is unity: $\sum_z P_z =
1$, where the sum runs over terminal vacua.  The connectedness of the
landscape then implies that all vacua have finite raw probability, if
there is at least one terminal vacuum.
\begin{figure}
\includegraphics[width=5cm]{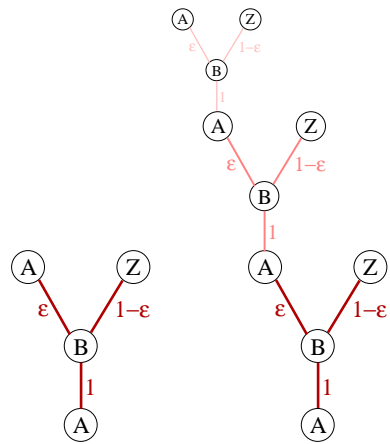}
\caption{\label{fig-abzpruned} Probabilities are easier to compute
  from the pruned tree, shown left for the ABZ model, with initial
  vacuum $A$.  One reads off readily that ${\cal P}_A = \epsilon$,
  ${\cal P}_B=1$, and ${\cal P}_Z=1-\epsilon$, which need only be
  normalized.  Right: The full tree can be recovered by iterating the
  pruned tree.  Each iteration changes all raw probabilities by the
  same factor, leaving the normalized probabilities invariant.}
\end{figure}

The {\em pruned tree\/} is constructed like the full tree, except that
one terminates the tree wherever it returns to the initial vacuum $o$
(Fig.~\ref{fig-abzpruned}).  One can compute raw probabilities by
applying Eq.~(\ref{eq-raw}) to the pruned tree.  Now the conservation
of the probability current implies that ${\cal P}_o + \sum_z {\cal
  P}_z = 1$, so that {\em all raw probabilities computed from the
  pruned tree are finite}.  Because $o$ is now effectively treated
like a terminal vacuum, this conclusion applies independently of the
presence of actual terminal vacua.

The full tree can be reconstructed from the pruned tree by joining a
copy of the pruned tree at the root to every final node labeled $o$ in
the original pruned tree, and iterating (see
Fig.~\ref{fig-abzpruned}).  This means that the raw probabilities of
the full tree, $P_i$, will be given by $P_i = {\cal P}_i (1 + {\cal
  P}_o + {\cal P}_o^2 + \ldots) = {\cal P}_i/\sum_z {\cal P}_z$. It
follows that the raw probabilities computed from the full tree, $P_i$,
converge if and only if the landscape is terminal.  If they do
converge, then the full and the pruned tree yield the same normalized
probabilities.  Thus, the pruned tree yields the most general
prescription.

\begin{figure}
\includegraphics[width=8.5cm]{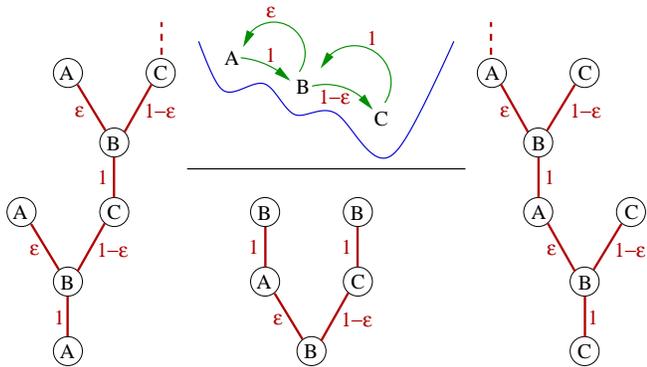}
\caption{\label{fig-abc} A landscape without terminal vacua.  For each
  initial vacuum, a pruned tree is shown.  For example, summation over
  paths in the left tree yields ${\cal P}_A = 1$, ${\cal P}_B =
  1/\epsilon$, ${\cal P}_C = (1-\epsilon)/\epsilon$.  After
  normalization, all pruned trees yield the same probabilities.}
\end{figure}

For example, consider a cyclic landscape with three metastable vacua,
as shown in Fig.~\ref{fig-abc}.  For simplicity, assume that $A$ and
$C$ cannot decay into each other directly but only through $B$.
The pruned tree depends on the initial vacuum, but the normalized
probabilities does not: $p_A = \epsilon/2$, $p_B = 1/2$, $p_C =
(1-\epsilon)/2$.  Below, it will be shown that $p_i$ is always
independent of the initial condition in a cyclic landscape.

\paragraph{Matrix formulation}
It is intuitive to compute the prior probabilities from tree graphs,
but is is also useful to reformulate the result as a matrix equation.
The {\em initial probability vector\/} ${\mathbf P}^{(0)}$, which
satisfies $P_j^{(0)}=1$ for $j=o$ and $P_j^{(0)}=0$ otherwise.  (The
result, Eq.~(\ref{eq-tm}) below, will naturally incorporate more
general initial probability distributions $\mathbf P^{(0)}$.)  Let us
consider the {\em partial probability\/} $P_i^{(\alpha)}$ to reach
vacuum $i$ from $o$ after exactly $\alpha$ steps.

On a full tree, the partial probabilities obey $\mathbf P^{(\alpha)} =
\eta \mathbf P^{(\alpha-1)}$.  The raw probability is the sum of
partial probabilities: $\mathbf P = \sum_{\alpha=1}^{\infty} \mathbf
P^{(\alpha)}$.  These two equations imply that the raw probability
obeys the matrix equation 
\begin{equation}
(1-\eta) \mathbf P = \eta \mathbf P^{(0)}~.
\label{eq-tm}
\end{equation}

To be consistent with the results above, this equation should have a
solution if and only if the landscape is terminal.  Let us prove this.
Suppose that there is no solution.  Then $(1-\eta)$ cannot be
invertible, and $\eta$ must have an eigenvalue $1$ with eigenvector
$\tilde \mathbf P$, satisfying $\tilde \mathbf P = \eta \tilde \mathbf
P$.  We are free to think of $\tilde \mathbf P$ as a partial
probability, in which case this describes an equilibrium: the
probability distribution is unchanged by an extra step.  By
connectedness of the landscape, this implies that the landscape
contains no terminal vacua.  Conversely, suppose that the landscape
has no terminal vacua.  This means that for any nontrivial initial
condition $\mathbf P^{(0)}\neq 0$, the vector $\eta \mathbf P^{(0)}$
must have some nonzero components, and in particular, their sum is
nonzero.  It also means that every column of $\eta$ adds up to $1$, so
the components of $(1-\eta) \mathbf P$ add to zero.  Thus,
Eq.~(\ref{eq-tm}) cannot be solved.

To deal with both the terminal and cyclic cases, I used pruned trees,
which are obtained by treating the vacuum $o$ as a terminal vacuum,
except in the first step.  Thus, pruned trees obey the matrix equation
\begin{equation}
(1-\eta S) \mathbf P = \eta \mathbf P^{(0)}~,
\label{eq-am}
\end{equation}
where $S$ annihilates the $o$-th column of $\eta$: $S_{ij} =
\delta_{ij} - P^{(0)}_i P^{(0)}_j$.  By the above proof, $(1-\eta S)$
is invertible so Eq.~(\ref{eq-am}) always has a unique solution.
Thus, it is the most general matrix equation we shall require.
However, in the terminal case, Eq.~(\ref{eq-tm}) is equivalent and
more elegant.

In fact, a simpler specialized equation is also available in the
cyclic case, since the pruned tree will have $P_o = P^{(0)}_o$ by
conservation of the probability current.  Hence $\eta S \mathbf P +
\eta \mathbf P^{(0)} = \eta (S \mathbf P + \mathbf P^{(0)}) = \eta
\mathbf P$.  Substitution into Eq.~(\ref{eq-am}) yields
\begin{equation}
(1-\eta) \mathbf P =0~.
\label{eq-nm}
\end{equation}

Note that $\mathbf P^{(0)}$ does not appear in Eq.~(\ref{eq-nm}).  This
demonstrates that the probabilities are independent of the initial
vacuum in the cyclic case, as advertised above.

To avoid confusion, let me emphasize once more that the general
(``pruned tree'') prescription is captured by Eq.~(\ref{eq-am}).  It
reduces to Eqs.~(\ref{eq-tm}) and (\ref{eq-nm}) for terminal and
cyclic landscapes, respectively.

In Ref.~\cite{VanVil06} the general prescription of Garriga {\em et
  al.}~\cite{GarSch05} was applied to the special case of cyclic
landscapes (for the simplest cyclic landscape the result was first
given in Ref.~\cite{BouFre06}).  The probabilities obey
Eq.~(\ref{eq-nm}), which shows that our prescription agrees with that
of Garriga {\em et al.}, if the landscape is cyclic.  For terminal
landscapes, such as the string landscape, the two prescriptions
differ.

False vacuum eternal inflation is particularly relevant to the
landscape, but it is straightforward to apply the above approach more
broadly.  If a worldline in slow-roll inflation (eternal or not) has
nonzero probability to end up in more than one vacuum, this can be
incorporated in the matrix $\eta$.  For example, the probability to
end up on either side of the asymmetric double well of
Ref.~\cite{GarVil01} is 50\%, if the worldline starts at the top of
the barrier; volume expansion factors do not enter.  If a vacuum has
continuous moduli, one can treat is as a (nearly) continuous set of
different vacua.

\paragraph{Weighting}
Having defined prior probabilities, let us now ask with what
probability $w_i$ observers will emerge in vacuum $i$.  The total
probability for $i$ to be observed is $p_i w_i/\sum_j p_j w_j$.

Anthropic arguments make sense only in a large and varied universe,
where they select for location (rather than for initial conditions, or
worse, for parameters of a fundamental theory).  With a much larger
cosmological constant but all other physics fixed, for example, it is
plausible that life would not have formed in our part of the
universe~\cite{Wei87}.

The problem is that other parameters are far from fixed in any
realistic landscape.  This poses a hard optimization problem,
requiring variations of the possible inflaton
potentials~\cite{TegRee97,FelHal05}, particle and force
content~\cite{HarKri06}, coupling constants, and other parameters.
Moreover, the challenge of identifying conditions for ``life'' will be
magnified, if the landscape contains low energy theories so different
from our own that we have little intuition for their impact on
cosmology or condensed matter physics.

But whatever observers may consist of, they must obey the laws of
causality, thermodynamics, and information theory.  Observers compute;
they store and retrieve information.  Because the causal diamond is
finite, the holographic approach makes it possible to quantify this
connection: the more free energy, the more likely it is that observers
will emerge.  More precisely, the number of possible operations should
be related to the free energy divided by the temperature at which it
is burned up.  This quantity is simply the increase in entropy.  Thus,
I propose to weight vacua by their entropy {\em difference},
$w_i=\Delta S(i)$, defined as the entropy leaving through the top cone
of the diamond minus the entropy entering the bottom cone of the
diamond.

It is important to stress what this weight does {\em not\/} depend on.
From a global viewpoint, it may seem natural that inflationary volume
factors (which are well-defined in noneternal slow-roll) should enter
directly into either the $p_i$ or the $w_i$~\cite{GarSch05}.  This
leads to a paradox~\cite{FelHal05}: the density perturbations should
be at an extreme end of the anthropically allowed window.  But from a
holographic point of view, volume produced in excess of one causally
connected region does not boost the likelihood of a vacuum further.
Inflation is useful in that it delays curvature domination, allowing
more free energy to be harvested; to this extent, it will enter
$\Delta S(i)$.  But there is no benefit in delaying it longer than
$|\Lambda|^{-1/2}$, the time when the cosmological constant begins to
dominate.

Similarly, one may be tempted to include the lifetime of a metastable
vacuum in its weight.  But stability matters only up to a point.  If
the decay disrupts the harvesting of free energy, it will enter the
weight factor $w_i=\Delta S$.  However, lifetimes can be exponentially
longer than the thermalization timescale; this does nothing to boost
the probability of observers.

The entropy production in our vacuum can be estimated, and its
dependence on various parameters yields constraints analogous to
anthropic bounds.  Unlike the latter, however, the weight $\Delta
S(i)$ can plausibly be computed also for distant regions of the
landscape~\cite{BouHar06}, at least when averaged over many vacua.
The entropy increase cannot be larger than the final entropy, which is
bounded in terms of the maximal area on the future boundary of the
causal diamond~\cite{Bou00a}.  For de~Sitter vacua this bound is
$3\pi/\Lambda$ .  In this sense, a small cosmological constant is
better than a large one, even when other parameters scan.  (R.~Harnik,
G.~Kribs, and G.~Perez have independently arrived at a similar
conclusion.)  This preference is only power-law, not exponential as a
purely statistical argument would imply.

Our vacuum has a positive cosmological constant, so its weight is
bounded.  Suppose that the landscape were infinite, in the sense that
parameters could scan arbitrarily dense discretua.  Then why don't we
find ourselves in a region that allows for even greater complexity
than our own?  The landscape must be finite, and numbers such as
$10^{-123}=e^{-283.2}$~\cite{Pol06} may turn out to be data points
that will help us determine its size empirically.

\paragraph{Acknowledgments} 
I would like to thank B.~Freivogel and J.~Polchinski for discussions.
This work was supported by the Berkeley Center for Theoretical
Physics, by a CAREER grant of the National Science Foundation, and by
DOE grant DE-AC03-76SF00098.

\bibliographystyle{apsrev}
\bibliography{all}
\end{document}